\pdfoutput=1
\documentclass[aps,nopreprint,twocolumn,prl,superscriptaddress,showpacs]{revtex4-1}
\usepackage{graphicx}
\usepackage{hyperref}
\hypersetup{%
 pdftitle = {Slit-array transmission loss feasibility in airborne sound},
 pdfauthor = {H. Estrada},
 bookmarksopen = true,
 colorlinks  = true,
 citecolor   = black,
 linkcolor   = black,
 raiselinks  = true,
 urlcolor    = blue,
 pdfstartview= FitH,
}

\begin{document}

\title{Slit-array transmission loss feasibility in airborne sound}

\author{H\'ector Estrada}
\email[To whom correspondence should be addressed.]{\ hecesbel@doctor.upv.es}
\affiliation{Centro de Tecnolog\'ias F\'isicas, Unidad Asociada ICMM- CSIC/UPV,
Universidad Polit\'ecnica de Valencia, Av. de los Naranjos s/n. 46022 Valencia,
Spain}
\affiliation{Instituto de Ciencia de Materiales de Madrid (CSIC), Cantoblanco,
28049 Madrid, Spain}

\author{Jos\'e Mar\'ia Bravo}
\affiliation{Centro de Tecnolog\'ias F\'isicas, Unidad Asociada ICMM- CSIC/UPV,
Universidad Polit\'ecnica de Valencia, Av. de los Naranjos s/n. 46022 Valencia,
Spain}

\author{Francisco Meseguer}
\affiliation{Centro de Tecnolog\'ias F\'isicas, Unidad Asociada ICMM- CSIC/UPV,
Universidad Polit\'ecnica de Valencia, Av. de los Naranjos s/n. 46022 Valencia,
Spain}
\affiliation{Instituto de Ciencia de Materiales de Madrid (CSIC), Cantoblanco,
28049 Madrid, Spain}

\date{\today}

\begin{abstract}
Recent experiments conducted in water at ultrasonic frequencies showed the
possibility of overcoming the transmission loss provided by homogeneous plates
at
certain frequencies by drilling periodically distributed holes on it. In this
letter, the feasibility of using slit arrays to increase the
transmission loss at certain frequencies for airborne sound is studied. Numerical results
predict a) very low transmission loss for a slit array in comparison with a
homogeneous plate in air and b) the transmission loss of a slit array can
overcome that of a homogeneous plate if the impedance mismatch is low enough.
\end{abstract}

\maketitle

In recent years, the study of the acoustical properties of hole and slit arrays
in finite plates has been boosted by the promising and interesting findings in
the field of optics \cite{genet2007,zheludev2008} and the development of
metamaterials for both, acoustic \cite{martinez-sala1995,sigalas2005} and
electromagnetic waves \cite{photcryst}. However, the use of hole and
slit arrays in the field of acoustics is not new. The main difference between
the already well known acoustic properties of slit or hole arrays
\cite{ingard1951,maa1998} and the latest research on such structures
\cite{zhangx2005,hou2007,lu2007,zhou2007,christensen2008,estrada2008b} lies in
the size of the wavelength with regard to the periodicity, aperture size, and
plate thickness. Although for light it is very interesting to increase the
transmission through metal plates by inserting subwavelength holes arrays, for
sound is always challenging to decrease the transmission. For finite thickness
plates or walls in the
long wavelength regime, the transmission is controlled by the mass per unit
area of the plate or wall \cite{FahySSVib}. Thus, sound screening at long
wavelength values is difficult to achieve in thin plates. Towards this direction
different strategies have been considered for hole arrays
\cite{estrada2008b,liu2008,zliu2010,christensen2010b}. 
Wood-anomaly sound screening is reported in \cite{estrada2008b} for periodically
perforated Aluminum plates immersed in water at ultrasonic frequencies. On the
other hand, two layers of periodically perforated plates have been theoretically
proposed to reduce the sound transmission  for low impedance mismatch
\cite{liu2008} (water-PMMA) and for infinite impedance mismatch
\cite{zliu2010,christensen2010b}
(rigid solid). 

One key issue in hole-array sound screening is whether it can be applied for the
case of airborne sound at audible frequencies
\cite{newsci2008,physworld2008,sciam2008}. In this paper we will show that due
to the high
impedance contrast at the solid/air interface, Wood-anomaly induced sound
screening of perforated plates does not overcome the high transmission loss
attained for the case of homogeneous plates. Numerical simulations also show
that transmission loss for the slit arrays, are effective only when the
impedance
mismatch is low, as in the case of Aluminum and water \cite{estrada2008b}. 

\begin{figure}
 \includegraphics[width=\columnwidth]{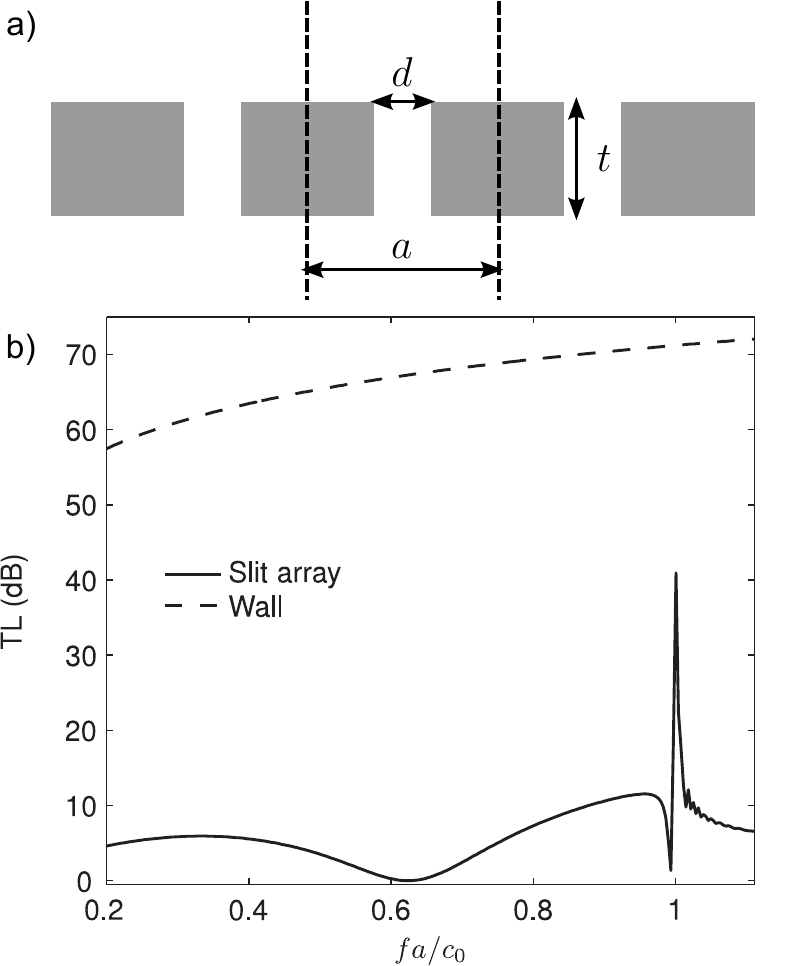}
\caption{a) Diagram of the slit-array geometry having a periodicity $a$, a slit
size $d$, and a slab thickness $t$. The unit cell is delimited by vertical
dashed lines. b) Transmission loss in dB for a slit array (solid curve) and a
homogeneous wall (dashed curve) both made of concrete in air as a function of
the normalized frequency at normal incidence.}
\end{figure} 

The transmission problem has been solved numerically by means of finite elements
implemented in Comsol Multyphysics software for frequency domain. A
unit cell of the slit array (see Fig. 1a)) having a period $a$, a slab thickness
$t=0.6a$, and an aperture of size $d=0.28a$ constitutes the geometry of the
problem. The slab is modeled as an elastic domain having zero out-of-plane
components of the strain and displacement to keep the problem in two dimensions.
A fluid domain is used to model the slit and the surrounding media. The pressure
in the fluid is decomposed as the sum of incident (known) and scattered
(unknown) pressures, the first being a plane wave at normal incidence. The
fluid-structure interaction is ensured by imposing continuity of both, the
normal displacement and the normal stress at the fluid-solid interfaces. The
periodicity enters through the lateral limits of the unit-cell via periodic
boundary conditions. Finally, to satisfy the Sommerfeld radiation condition at
infinity, perfectly matched layers (PML) are used at top and the bottom of the
unit-cell. Wavelength-dependent scaling is applied to the mesh, the thickness of
the PML, and the vertical size of the unit-cell. The transmitted and reflected
sound power is calculated by integrating the vertical sound intensity along the
unit-cell width right at the interface between the fluid domain and the PML.
Convergence is achieved for a mesh element size around $\lambda/15$ and has been
tested through the balance of the total sound power. 

The characteristic acoustic impedance $z_0$ is given by the product between the
fluid density $\rho_0$ and the phase velocity $c_0$. The impedance mismatch
between a solid and a fluid will be simple considered as $K=z_s/z_0=\rho
c_l/\rho_0 c_0$, where $\rho$ is the solid density and $c_l$ is the longitudinal
wave velocity in the solid. In the case of concrete ($\rho=2400$ kg$/$m$^3$,
$c_l=2996$ m$/$s) and air ($\rho_0=1.12$ kg$/$m$^3$, $c_0=343$ m$/$s) the
impedance mismatch yields $K=1.7\times 10^4$. The results of this high impedance
mismatch in the transmission loss (TL) of the slit array compared with the
homogeneous wall of the same thickness  and material is showed in Fig. 1(b) as a
function of the normalized frequency $fa/c_0$ at normal incidence. Resonant full
transmission peaks and the Wood anomaly are present in the slit array spectrum,
as expected from previous results using rigid-solid assumption
\cite{zhangx2005,hou2007,lu2007,zhou2007,christensen2008,estrada2008b}. It is
clear that the slit array doesn't provide any advantage at any frequency in
terms of TL over the homogeneous plate. The homogeneous
plate TL (dark solid curve) is nearly 55 dB below that for concrete-air (Fig.
1b)). For simplicity, one can define the slit array insertion loss as
IL$=10\log(\tau/\tau_0)$, where $\tau$ is the transmitted sound power
coefficient of the slit-array for finite impedance mismatch and $\tau_0$ is the
transmitted sound power coefficient of the homogeneous layer having the same
material and thickness than the slit array. One can see the insertion loss (IL)
of the
slits in a homogeneous plate will be everywhere negative. Changing the fluid to
water ($\rho_0=1000$ kg$/$m$^3$, $c_0=1480$ m$/$s) and the solid to Aluminum
($\rho=2700$ kg$/$m$^3$, $c_l=6500$ m$/$s) lowers the impedance mismatch up to
$K=1.2\times 10^1$, which is 3 orders of magnitude lower than the previous case.
The TL for Al-water slit array and homogeneous plate as a function of the
normalized frequency at normal incidence is showed in Fig. 2. 
Between $0.7<fa/c_0<0.98$, the slit array TL overcomes the plate TL
reaching a maximum of 21 dB. The differences in the slit array transmission for
concrete in air and Al in water can be extracted comparing Fig.1(b) and Fig. 2.
The first resonant full transmission peak $fa/c_0 \approx 0.6$ is slightly
shifted to lower frequencies with regard to the concrete-air case, while the
Wood anomaly minima is almost absent for the Al-water slit array. The TL maximum
at $fa/c_0\approx0.9$ is almost 10 dB larger for the Al-water array compared
to the concrete-air array. This results explain the results reported in
\cite{estrada2008b} for Al-water mismatch at ultrasonic frequencies, although
hole arrays where studied instead of slit arrays.

\begin{figure}
 \includegraphics[width=\columnwidth]{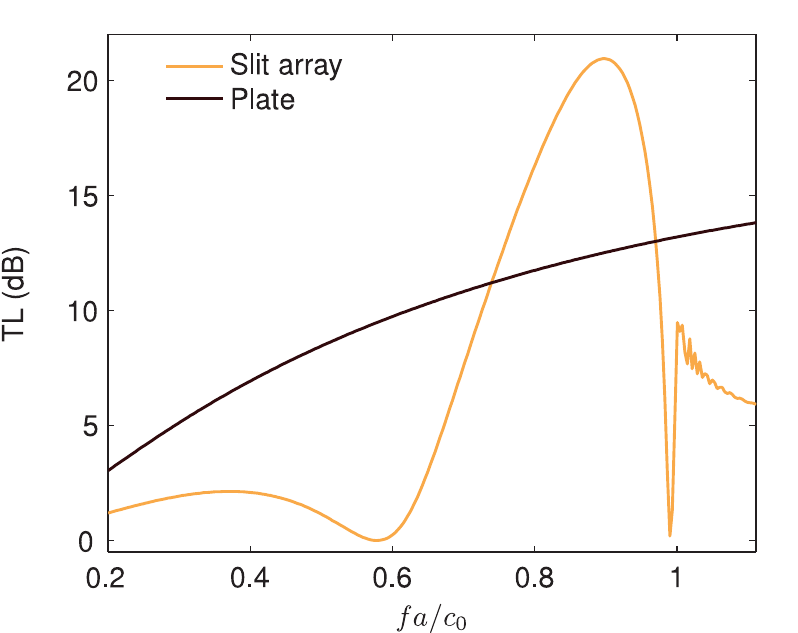} 
\caption{Transmission loss in dB of a slit array (light solid curve) and a
homogeneous plate (dark solid curve) both made of Aluminum in water as a
function of the normalized frequency at normal incidence.}
\end{figure} 

\begin{figure}
 \includegraphics[width=\columnwidth]{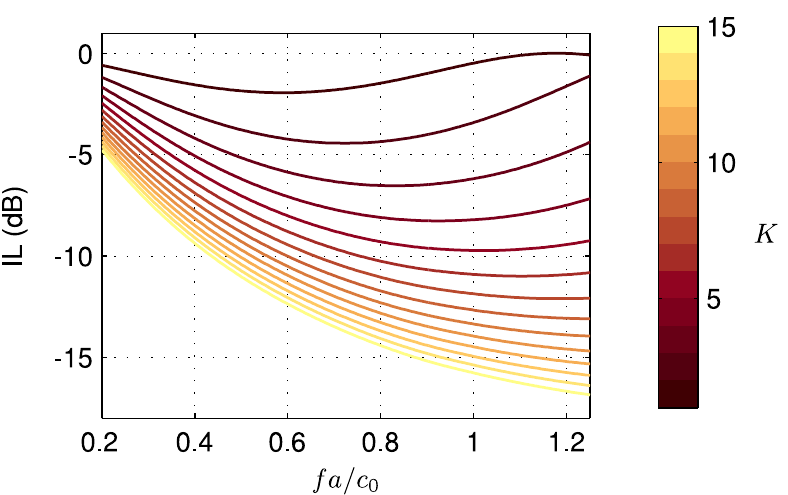}
\caption{Insertion loss in dB of slit arrays in a fluid layer as a function of
the normalized frequency for the different values of $K$.}
\end{figure} 

The previous results preclude the use of the slit array sound screening in
airborne conditions due to the huge impedance mismatch between the air and most
solids ($K> 10^3$). In order to establish a proof-of-concept we performed
further calculations replacing the solid by a fluid having different acoustic
impedances. 
The results of the IL in dB for
different $K$ are shown in Fig. 3 and set a clear distinction between the
phenomena observed with solid slabs and fluid slabs. An homogeneous fluid slab
can only sustain Fabry-Perot modes and in all cases the TL provided by the
homogeneous slab is several decibels higher than that of the slit array.
Therefore, the IL is almost always negative and decreases as $K$ increases.
There are, however,  porous materials capable of showing low impedance and
certain amount of sound absorption as well. This kind of materials could be
suitable candidates to show interesting properties when they are arranged
periodically, as several authors have reported in previous studies for airborne
sound \cite{takahashi1989,mechel1990}. Porous materials are, however, more
complex and are out of the scope of this study.  
The possibility of overcoming the homogeneous plate transmission loss by
inserting slits in a low impedance material is unfortunately not realistic for
airborne sound because no conventional solid has such a low acoustic impedance.
Replacing the solid by a fluid , i.e. another gas, could be more feasible in the
practice but, as our calculation predicts, will be useless in terms of
transmission loss.
Also low impedance metamaterials \cite{torrent2007} working in the effective
media regime could be appropriate to provide the low impedance mismatch
required. We hope this study stimulates more research on slit arrays made of
low-impedance porous materials or metamaterials. 

\begin{acknowledgments}
The authors wish to acknowledge financial support from projects MICINN MAT2010-16879, 
Consolider Nanolight.es CSD-2007-0046 of the Spanish Education and 
Science Ministry, and project PROMETEO/2010/043 of Generalitat Valenciana. H. E.
acknowledges a CSIC-JAE scholarship.
\end{acknowledgments}


%

\end{document}